# Enabling Gravity-Physics by Inquiry using Easy Java Simulation

Loo Kang WEE[1], Giam Hwee GOH[2] and Charles CHEW[3]

[1]Ministry of Education, Education Technology Division, Singapore
[2]Ministry of Education, Yishun Junior College, Singapore
[3]Ministry of Education, Academy of Singapore Teachers, Singapore

wee_loo_kang@moe.gov.sg, goh_giam_hwee@moe.edu.sg, charles_chew@moe.gov.sg

Abstract: Studying physics of very large scale like the solar system is difficult in real life, using telescope on clear skies over years. We are probably a world-first to create four well-designed gravity computer models to serve as powerful pedagogical tools for students' active inquiry, based on real data. These models are syllabus-customized, free and rapidly-prototyped with Open-Source-Physics researchers-educators. Pilot study suggests students' enactment of investigative learning like scientist is now possible, where gravity-physics 'comes alive'. We are still continually improving the features of these computer models through feedback from students and teachers and the models can be downloaded from http://weelookang.blogspot.sg/2012/01/gravity-physics-by-inquiry-2012-innergy.html. We hope more teachers will find the simulations useful in their own classes and further customized them so that others will find them more intelligible and contribute back to the wider educational fraternity to benefit all humankind.

PACS: 01.50.H-  91.10.-v  91.10.Sp  96.20.Jz  04.80.-y  96.20.Jz

## I. INTRODUCTION

Imagine sending students into outer space to collect gravitational scientific data and visualize the planets in the solar system (see Figure 1), or be in outer space just outside Earth's atmosphere to visualize geostationary satellites (Wee & Goh, 2013) (see Figure 2). How about science laboratory toolkit that allows students to investigate the gravitational effects of isolated mass that cannot be observe on Earth (see Figure 3) or visit the Earth's Moon to launch a rocket out into space to investigate what is the minimum kinetic energy required to escape the Moon's and Earth's gravity pull (see Figure 4)?

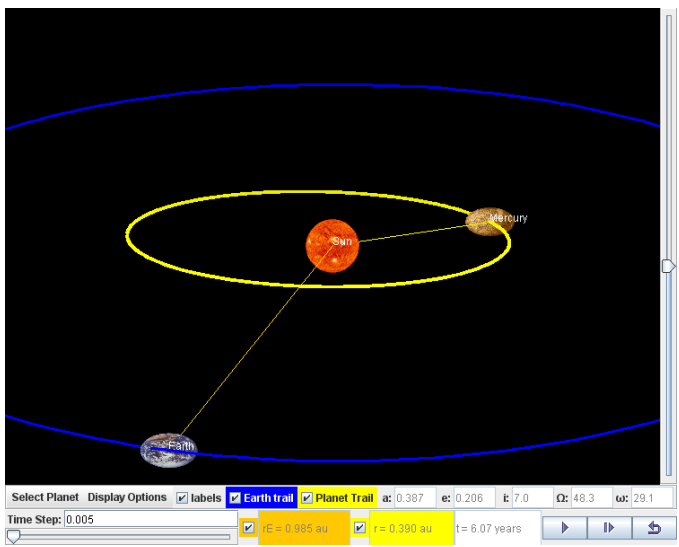

Figure 1. Kepler Solar System Model (Timberlake & Wee, 2011) with actual astronomical data built into the simulation, with realistic 3D visualization, (radius of planets such as Earth, rE and another planet for comparison r, and time t for determination of period of motion, T) data for inquiry learning and to situate understanding

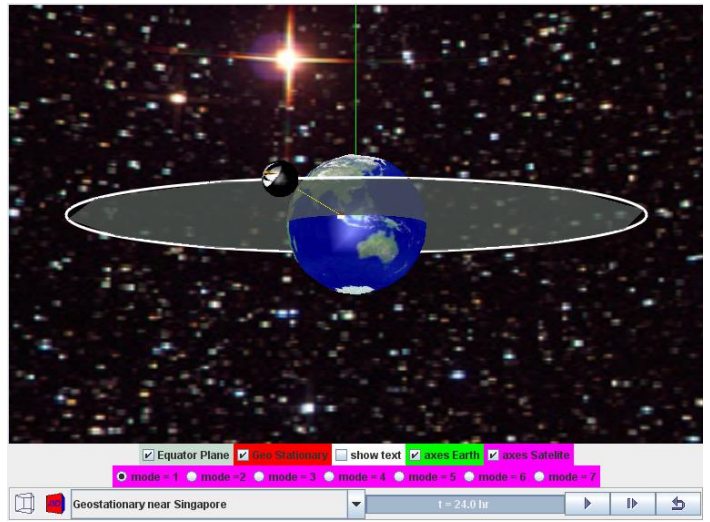

Figure 2. Geostationary Satellite around Earth Model (Wee & Esquembre, 2010) suitable for inquiry learning through different mode =1 to 7, with Geo Stationary checkbox option, 3D visualization, customized with Singapore as a location position for satellite fixed about a position above the earth with period 24 hours, same rotation sense on the equator plane.

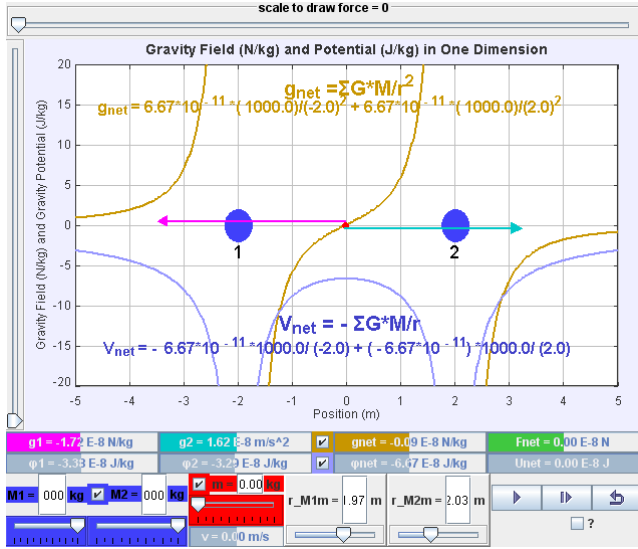

Figure 3.  One Dimensional Gravitational Model (Duffy & Wee, 2010a) suitable for investigative inquiry learning through data collection, customized with syllabus learning objectives such as gravitational strength g, gravitational potential φ when one or both masses M1 and M2 are present with a test mass m. Superimpose are the mathematical representations, vector presentation of g, based on current Newtonian model of gravity.

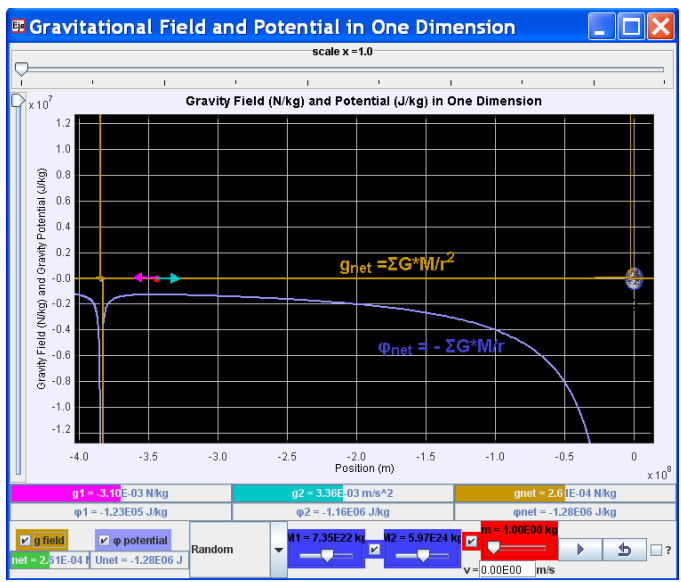

Figure 4.  One Dimension Gravitational Moon-Earth Model (Duffy & Wee, 2010b) suitable for investigative inquiry learning, further customized to allow the experiencing of an Advanced Level examination question June 87 /II/8. Data are based on real values where students can play and experience physics otherwise difficult to related to examination question.

It would be a great financial burden to space-shuttle classroom full of students into outer-space and not forgetting a potentially dangerous journey without oxygen and in extreme low temperatures. Thus, we believe that there is justification to 'bring' the planets in the solar system and other outer space environments into a typical classroom and put the students in a position to conduct virtual experiments using teacher-researcher created computer models (Psycharis & Aspaite, 2008), or in short, simulations.

Thus, building on open source codes shared by the Open Source Physics (OSP) community, we customized these four computer models, also known as simulation, to allow our students to role play as scientists interacting with the inquiry-designed physics-equations-modeled phenomena. This is done using a free authoring toolkit called Easy Java Simulation (EJS) (Esquembre, 2012).

These computer models can be downloaded from http://weelookang.blogspot.sg/2012/01/gravity-physics-by-inquiry-2012-innergy.html, digital libraries in ComPadre Open Source Physics and NTNUJAVA Virtual Physics Laboratory, creative commons attribution licensed.

The recommended system requirement for running EJS models is the Intel Pentium processor.

## II.  RESEARCH

The process to conceptualize and develop the innovation started in 2007 when we found the Open Source Physics (OSP) community and through our teacher leadership (MOE, 2009), aiming to bring world class OSP computer models into Singapore and the world's classrooms.

We immersed in discussions forums mainly on NTNU Java Virtual Lab (Hwang, 2010) and OSP (Christian, 2010) and networked learnt in these communities (Hord, 2009). This was how we initiated teacher-led process in networked learning with the world's best computational physicists as in Figure 5 since 2007 and lesson intervention in 2011.

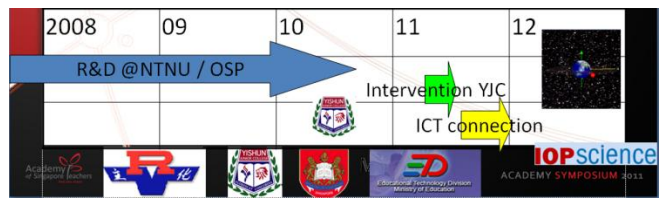

Figure 5.  Simplified timeline showing research and development at NTNU Java Virtual Lab (Hwang, 2010) and OSP (Christian, 2010) in 2008 to 2010 and intervention and sharing on ICT connection edumall in 2011, research journal publication at Institute of Physics - Physics Education planned 2012

We also submit our Digital computer models to the Open Source Physics Library (peer-reviewed by Physics Professors) based in USA as well as publishing journal papers in Institute of Physics (IOP) Physics Education journal (see Figure 5) based in Europe, ensuring research rigor and acceptance by the Physics research community.

Furthermore, educational research and computer models from the OSP community provided suitable 'templates' for our computer models to be derived or remixed from. This, in some way, provided 'guaranteed' scientific validity in our models.

We are pleased to report that OSP recently received the Science Prize for Online Resources in Education (SPORE) Prize (Christian, Esquembre, & Barbato, 2011) honored by Science Magazine established to encourage innovation and excellence in education, in the use of high-quality on-line resources by students, teachers, and the public in the world. This is a piece of good news for us too as we are active contributor(s) to the OSP digital library since 2009 with 10 out of the 550 computer models/resources shared world-wide through the OSP website for free, benefiting humankind regardless of race, language or religion.

## III.  FOUR COMPUTER MODEL DESIGN IDEAS

To add to the body of knowledge surrounding why simulations could be effective pedagogical tools (Wee, Chew, Goh, Tan, & Lee, 2012), we share four computer model design ideas-insights (Wee, 2012b) that we believed have raised the effectiveness and usefulness of the tool for students' active learning.

*A. Kepler's System Model*

Kepler's System Model (Timberlake, 2010) by Professor of Physics, Berry College, USA served as the template for our Kepler's Solar System Model (Timberlake & Wee, 2011) (see Figure 6).

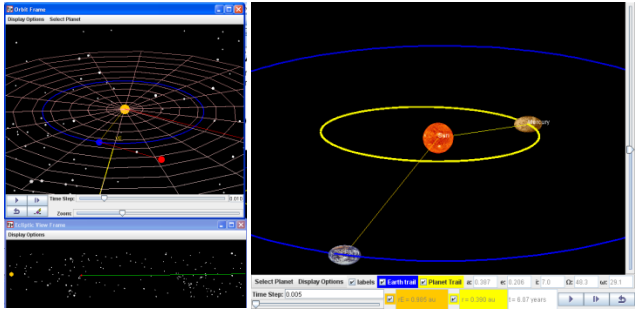

Figure 6.  Kepler System Model (Timberlake, 2010) (left) and our customized model (Timberlake & Wee, 2011) (right) notice our model is focused and can simulate all planets moving at the same time, better graphics of the planets etc.

*1) Enhanced realism for associated learning*

We added Uranus, Neptune and Pluto using real life astronomy data from NASA website to allow students to collect data on these planets. Realistic pictures of planets were also added for faster recognition and better associated learning. The source code was also re-programmed such that all planets will move together instead of only 3 planets as in (Timberlake, 2010).

*2) Simplified view and data collection*

We simplified the 3D world view showing only the planets and their respective trails so that students can take measurement of the time of each planet to travel a complete revolution around the Sun. Data like radius of orbits and time of orbits is now clearly noticeable for self-directed inquiry.

*B. Geostationary Satellite around Earth Model*

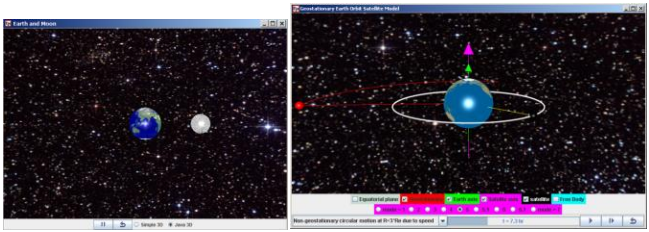

Figure 7.  Earth and Moon Model (Esquembre, 2010) (left) and our customized model (Wee & Esquembre, 2010) (right), notice the customization created to suit our own learning and teaching objectives, with references made to geographic location of Singapore.

*1) Simple and realistic 3D view and associated learning to real world*

A realistic Java 3D implementation view allows visualization of Earth to scale drawing of a scale of $1 \times 10^6$ m with $R_{Earth} \approx 0.637$ and radial distance from centre of Earth to geostationary orbit, $R \approx 4.23$ . To strengthen the associated learning for Singapore students, we used three positions with land mass around the equator such as South East Asia (Singapore), Africa and America continent (Figure 8).

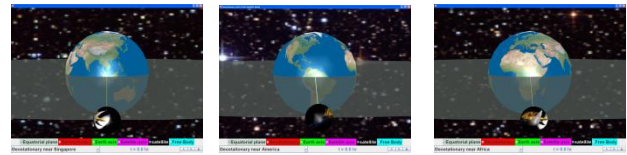

Figure 8.  Orbital view of Earth with geostationary satellite above South-East Asia (Singapore), Americas and Africa continents for associated learning to real world.

To aid in the visualization from different perspectives, a semi transparent equatorial plane and axis of rotations for the Earth and satellite were added in to help students gain a better visualization of the 3D space that they are interacting with.

*2) Incorrect physics for conceptual reasoning*

Incorrect physics such as a geostationary satellite above a point on the northern hemisphere of Earth was used to challenge thinking about what is 'wrong' with this orbit. A free body diagram showing the equal and opposite forces acting separately on the Earth and satellite helped students to use what they learnt about Newton's Third Law in this context.

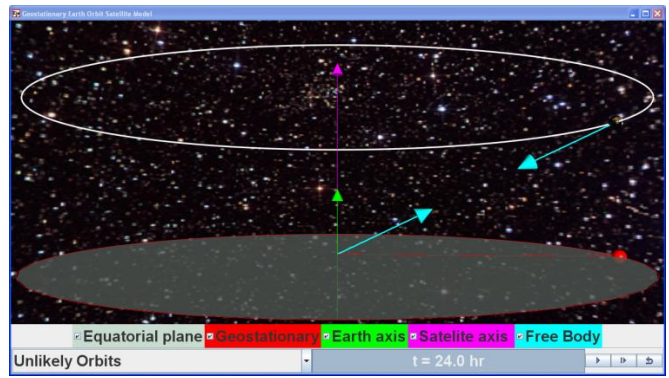

Figure 9.  A free body diagram of Earth and satelite showing the forces (teal) on the Earth and satellite as equal and opposite in direction acting on different bodies.

*C. One Dimensional Gravitational Model*

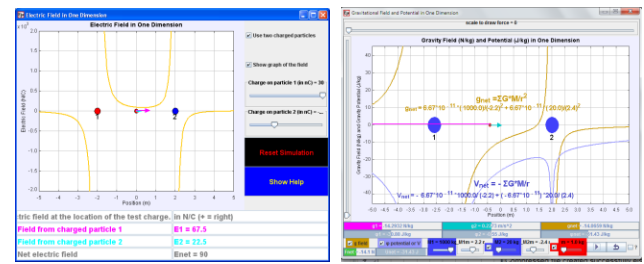

Figure 10. Point Charge Electric Field in 1D Model (Duffy, 2009) (left) and our customized model (Duffy & Wee, 2010a) (right) notice play button is previous not available and additional potential V or φ concept.

*1) Enhanced realism for associated learning*

Our quick literature review suggests we are probably a world first to create a 2-mass gravitational model with real life data associated with the universal gravitational constant. Andrew's original model was on electrically charged particles, while we customized it to gravitational mass particles in this model. Realism was also added via the customization-addition of motion of the test mass under the influence of the 2-mass system.

*2) Visualization of invisible concepts*

We added the forces acting on the test mass, gravitational field strength and potential due to the 2-mass system for improving the visualization of these invisible field effects. We speculate that the multiple scientific representations (Wong, Sng, Ng, & Wee, 2011) can be made more useful by showing the actual numerical data associated with the test mass position and in SI units instead of arbitrarily set values which is what makes our model relevant to the real world.

*D. One Dimensional Gravitational Model*

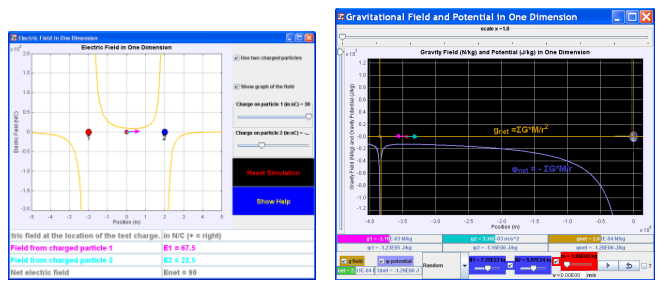

Figure 11. Point Charge Electric Field in 1D Model (Duffy, 2009) (left) and our customized model (Duffy & Wee, 2010b) (right) notice real astronomical data are programmed as that the values reflect actual numerical calculated from actual theoretical experiments.

*1) Enhanced realism for associated learning*

Pictures of Earth and Moon with actual astronomical mass data and the universal gravitational constant are designed into the model to represent the actual numerically calculated values of escape velocity from Earth's surface as 11,200 km/s. This strengthens students' confidence in the theoretically calculated values as the simulation reflects the physics they learn in the textbooks. This computer model serves to address the common inability of students to relate to an Advanced Level Physics examination question (June 87/II/8), bringing textbook context alive for meaningful play (Lee, 2012).

Thus, we have elaborated on why our research on these four computer models to create interactive engagement learning by standing on the shoulders of OSP giants, is a fundamental breakthrough research, never accomplished before in Singapore schools, at practically zero cost.

IV. FEEDBACK FROM STUDENTS

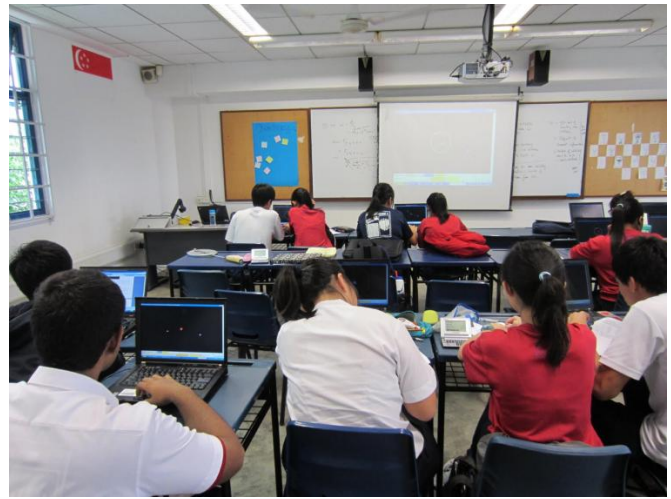

Figure 12. Typical classroom setup where students self direct the inquiry learning collaboratively or otherwise, using the computer models as referents (Dede, Salzman, Loftin, & Sprague, 1999), can served as powerful learning tools when well facilitated by teacher(s). Picture by Goh G.H.

We include excerpts from the qualitative survey results and informal interviews with the students to give some themes and insights into the conditions (Figure 12) and processes during the laboratory lessons. Words in brackets [ ] are added to improve the readability of the qualitative interviews.

*1) Improved visualization of invisible concepts and cannot be seen in real life*

"It make[s] the theory much more easier to understand, especially when it is difficult to conduct experiments to prove the Newton's law of gravitation and the Kepler's third law, for everything occurs in space".

"The [information and communication technology] ICT lessons make it easier for us to depict the motion of objects in a clearer manner and drawing the diagrams in questions easier. As some of the programmes possess 3-dimensional views, we are able to view the motion of the object in 3-dimensional, and hence, further explains the [worksheets] question with the use of ICT programmes."

"The lessons allow me to understand the movement of a satellite which we cannot see normally in real life and are unable to comprehend from the 2D [textbook] diagram. Thus, the 3D simulation allows me to learn better."

"Allow [me] to get a better understanding of the topic as simulation aids in visualizing the various questions easily, thus, able to solve the question. The lessons give me a clearer explanation of how things works thus, allowing me to understand".

*2) Enable Self directed inquiry learning of suitable pace*

"These lessons allow me to learn physics concepts better by using applications in the future. Thus, with the help of these applications and programs, I will be able to learn physics concepts through self-learning in future. Hence, it is good."

"I feel better as I am able to explore on my own."

*3) Learning with computer models can be fun and beyond syllabus*

"The lessons were quite a success as it enable[s] us to learn physics through another platform. It makes lesson more fun and interesting."

"It provided interesting information and insights on things outside of the syllabus."

*4) Appreciative learners*

"I would like to show my appreciation for the [information and communication technology] ICT inventors and teachers who participated in this ICT learning programme as it is a new opportunity for us to pick up high technology skills to pick up physics."

"I would like to thank my teacher for allowing us to gain exposure to these simulations and how they are able to be used to allow us [to] understand the topic better."

## V. CONCLUSION

We argue for computer models as suitable physics learning environments for the following three reasons: 1) to visualize physics through multiple representations especially for invisible and very large scale concepts 2) ease of theory generation from real-life, "annoyance-free" and accurate simulations, 3) mathematical analysis & modeling to deepen inquiry.

We demonstrate how our school-based research on these four computer models on gravity, which stands on the shoulders of the Open Source Physics giants, are research-validated. Global research community innovative process using free tool has created computer models to achieve student-directed gravity-physics by inquiry with simulations.

The free computer models can be downloaded from http://weelookang.blogspot.sg/2012/01/gravity-physics-by-inquiry-2012-innergy.html, ComPadre Open Source Physics (Wee, 2012a) and NTNU Virtual Physics Laboratory (Timberlake & Wee, 2011; Wee, Duffy, & Hwang, 2012a, 2012b; Wee & Esquembre, 2010) digital libraries.

General feedback from the students has been relatively positive, triangulated from the survey responses, interviews with students and discussions with teachers.

We hope more teachers will find these computer models useful and can act more intelligible (Juuti & Lavonen, 2006) in their own classes.

## ACKNOWLEDGEMENT

We wish to acknowledge the passionate contributions of Francisco Esquembre, Fu-Kwun Hwang and Wolfgang Christian for their ideas and insights in the co-creation of interactive simulation and curriculum materials.

This research is made possible; thanks to the eduLab project NRF2011-EDU001-EL001 Java Simulation Design for Teaching and Learning, awarded by the National Research Foundation, Singapore in collaboration with National Institute of Education, Singapore and the Ministry of Education (MOE), Singapore.

Lastly, we also thank MOE for the recognition of our research on the computer model lessons as a significant innovation with the 2012 on this MOE Innergy (HQ) GOLD Awards by Educational Technology Division and Academy of Singapore Teachers.

AUTHOR

| | |
|---|---|
| 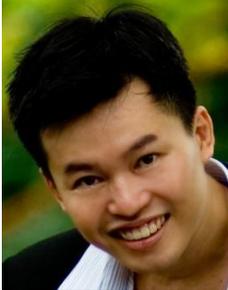 | Loo Kang Lawrence WEE is currently an educational technology specialist at the Ministry of Education, Singapore. He was a junior college physics lecturer and his research interest is in Open Source Physics tools like Easy Java Simulation for designing computer models and use of Tracker. |
| 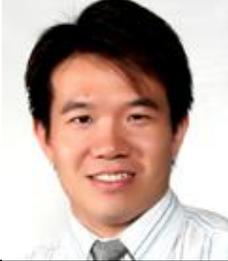 | Giam Hwee Jimmy GOH is currently the Head of Science Department in Yishun Junior College, Singapore. He teaches Physics to both year 1 and 2 students at the college and advocates inquiry-based science teaching and learning through effective and efficient means. |
| 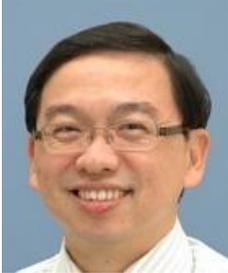 | Dr Charles CHEW is currently a Principal Master Teacher (Physics) with the Academy of Singapore Teachers. He has a wide range of teaching experiences and mentors many teachers in Singapore. He is an EXCO member of the Educational Research Association of Singapore (ERAS) and is active in research to strengthen theory-practice nexus for effective teaching and meaningful learning. |